# Spatiotemporal Analysis for Frequency–Magnitude Distribution of Earthquakes Across Mainland China: Comparing Classical b-value and b-positive Across Tectonic Regimes

**Yuxin Zhou [1,2], Huai Zhang [1*], S. Mostafa Mousavi [2]**

[1] State Key Laboratory of Earth System Numerical Modeling and Application, College of Earth and Planetary Sciences, University of Chinese Academy of Sciences, Beijing, China 100049.

[2] Department of Earth and Planetary Sciences, Harvard University, Cambridge, MA 02138, USA

[*] *Corresponding author: Huai Zhang (hzhang@ucas.ac.cn)*

**Key Points:**

- Spatially heterogeneous b-values in mainland China are controlled by local topography and fault regimes.
- The localized high b-value has been further validated as being consistent with regions previously interpreted as having fluid involvement.
- The b-positive method exhibits relative stability in handling catalog incompleteness and temporal heterogeneity.
- Expanding the existing seismicity analysis tool with a b-value spatial mapping module provides a more comprehensive framework for the community.

## Abstract

The Gutenberg-Richter power-law relationship essentially governs the frequency-magnitude distribution (FMD) of seismic activity. The b-value, as the slope of this distribution, is widely recognized not only as a quantifier of the relative proportion of small to large events but also as a diagnostic precursor of crustal differential stress and potential big to huge earthquakes. This study investigates the spatiotemporal evolution of b-values across major tectonic regions in mainland China, for instance, the Sichuan-Yunnan, North China, and Northwest regions. Our analysis of over 50 years of seismic records systematically examines b-value variations across major seismogenic zones in mainland China to evaluate their utility as precursors for moderate-to-strong earthquakes. Moreover, our comparison suggests that the b-positive estimator is less affected by changes in detection capability and short-term aftershock incompleteness than the classical b-value estimator in the cases examined. While the Sichuan-Yunnan region displays a complex "mosaic" of high and low b-values reflecting fluid-driven processes and localized locking, the North China Plain and Northwest regions exhibit consistently lower b-values, suggesting high background stress and rigid crustal integrity. We also find that several of the rupture areas of moderate-to-large earthquakes in our catalog coincide with localized low-b patches identified in earlier long-term windows. This study provides a foundation for systematic, prospective evaluation of b-value-based precursor monitoring in mainland China.

## Plain Language Summary

The frequency of earthquakes follows a predictable pattern: small earthquakes occur far more often than large ones. The slope of this relationship, known as the b-value, can reveal how much stress is building up in Earth's crust — lower b-values generally indicate higher stress and greater potential for large earthquakes. In this study, we analyzed more than 50 years of earthquake records across three major seismic regions in mainland China: Sichuan-Yunnan, North China, and Northwest China. We tracked how b-values changed over time and space to identify zones that may be at elevated risk for future large earthquakes. We also compared the traditional b-value method with a newer, more robust approach, the b-positive method, which performs better when earthquake catalogs are incomplete or of uneven quality. Our results show that the Sichuan-Yunnan region exhibits a patchy distribution of high and low b-values, likely reflecting the influence of underground fluids and localized fault locking. In contrast, North and Northwest China show persistently low b-values, indicating high crustal stress and increased seismic hazard. These findings highlight how mapping b-value anomalies can help identify high-risk fault segments, offering a practical tool for improved earthquake hazard assessment in tectonically complex regions.

## 1 Introduction

The frequency-magnitude distribution (FMD) of seismic activity statistically follows the Gutenberg-Richter (G-R) power-law relationship (Gutenberg & Richter, 1944) empirically expressed as:

$$\log_{10} N = a - bM_c \tag{1}$$

Here， $N$ represents the cumulative frequency of earthquakes with a magnitude greater than or equal to the completeness of magnitude Mc. $a$ describes the seismicity rate (i.e., the background earthquake rate) in the study area. At the same time, b is the slope of the power-law distribution and quantifies the relative proportion of small to large earthquakes (Schorlemmer et al., 2005).

It has been argued that b-value may carry diagnostic information about the state of stress in the faults that generate earthquakes (Amitrano, 2003; Gulia et al., 2018). The existing experimental and observational studies widely support a negative correlation between the b-value and crustal differential stress: high-stress locked regions often exhibit low b-value anomalies, while stress-release zones or highly fractured areas correspond to higher b-value (Narteau et al., 2009; Ogata et al., 1991; Ogata & Katsura, 1993; Scholz, 1968). Based on this physical mechanism, the b-value has become a core parameter for estimating recurrence intervals of different fault types and optimizing probabilistic seismic hazard assessments.

Beyond the traditional stress-dependence model, recent laboratory studies emphasize the critical role of fault-zone geometry and surface roughness in modulating b-values (Piegari et al., 2025). Laboratory stick-slip experiments (T. H. W. Goebel et al., 2017) demonstrate a positive correlation between fault roughness and b-values: rougher faults with abundant small-scale asperities produce more distributed seismicity (high b), whereas smoother, more mature faults favor larger ruptures (low b).

Over the past two or three decades, the b-value has gained widespread recognition as an effective tool for investigating seismicity patterns (Chen et al., 2006; A. Tiwari et al., 2023; R. K. Tiwari & Paudyal, 2023). It has been suggested that changes in b-value after an earthquake can be used to discriminate whether that earthquake is part of a foreshock sequence or a more typical mainshock-aftershock sequence, with a decrease in b-value heralding a larger earthquake to come. Despite these promising applications, the use of b-values for precursor identification is fraught with statistical pitfalls. However, the use of b-values for precursor identification has substantial statistical pitfalls: many reported pre-event 'anomalies' may be artifacts of catalog incompleteness, small sample sizes in sliding windows, or random fluctuations that mimic precursory signals, leading to high false-alarm rates (Kamer & Hiemer, 2015).

To deepen understanding of its spatiotemporal evolution, researchers have been dedicated to precisely calculating the b-value and its associated uncertainty. Currently, the classic algorithm for estimating the b-value and its standard error is the Maximum Likelihood Estimation (MLE) method (Aki, 1965; Bender, 1983; Shaw et al., 2025; Shi & Bolt, 1982; Utsu, 1965).

In China, seismological research primarily focuses on Southwest China, where tectonic deformation is most intense. Within this region, the majority of studies concentrate on the Longmenshan area (Deng et al., 2010), the seismogenic zone of the 2008 Ms 8.0 Wenchuan earthquake (Jia et al., 2012; Liang et al., 2025; S. Zhang & Zhou, 2016). Additionally, some studies have focused on seismic catalog analysis in specific regions such as Gansu (H. Wang et al., 2025), Yunnan (Y. Zhang et al., 2025), and the northern part of the North China Craton (Bi et al., 2025).

Studies that span broad spatial scales often overlook the inherent discrepancies across magnitude scales (e.g., ML, Ms, Mw), which introduce substantial uncertainty into Mc and b-value estimates. We address this by applying a magnitude-homogenization procedure to the entire CENC catalog, projecting all events onto a common Mw scale.

On the other hand, van der Elst (2021) argues that traditional methods are flawed because earthquake catalogs are often incomplete immediately after a large event, thereby biasing the data and leading to false conclusions about stress changes. They introduced b-positive, which is a new statistical method for measuring the b-value during active earthquake sequences, particularly aftershocks, defined as:

$$\widetilde{\beta}' = \left(\overline{|m'|} - M_c'\right)^{-1} \tag{2}$$

Here, $\overline{|m'|}$ is again the sample mean of the absolute magnitude differences, and $M_c'$ is a minimum magnitude difference. While the b-positive approach theoretically offers greater stability than the classical b-value, the latter remains the industry standard. Consequently, the universality and reliability of this new method necessitate further validation across a broader range of practical applications.

Motivated by these considerations, this study presents a systematic, mainland-wide retrospective analysis of b-value variations in China. We apply a magnitude-homogenization procedure to the 1970–2024 CENC catalog and partition the analysis spatially and temporally to estimate completeness locally. We compute both the classical and b-positive estimators in parallel, evaluate their stability through a multi-parameter sensitivity test, and assess the statistical significance of all reported variations using the b-significant autocorrelation method (Mirwald et al., 2024). To support these analyses, we extended the SeismoStats package (Mirwald et al. 2025) with a b-value spatial-mapping module, which is released as an open-source toolkit. The resulting baseline maps and time series provide a community reference for b-value behavior across distinct tectonic regimes of mainland China and a head-to-head comparison of the two estimators under continental catalog conditions. Furthermore, this work highlights the potential of b-value in seismic hazard assessment and emergency response, while validating the practical utility of the b-positive method within complex tectonic environments.

## 2 Materials and Methods

This study employs the earthquake catalog provided by the China Earthquake Networks Center (CENC). The dataset covers the period from 1970 to 2024 in mainland China (Fig. 1a).

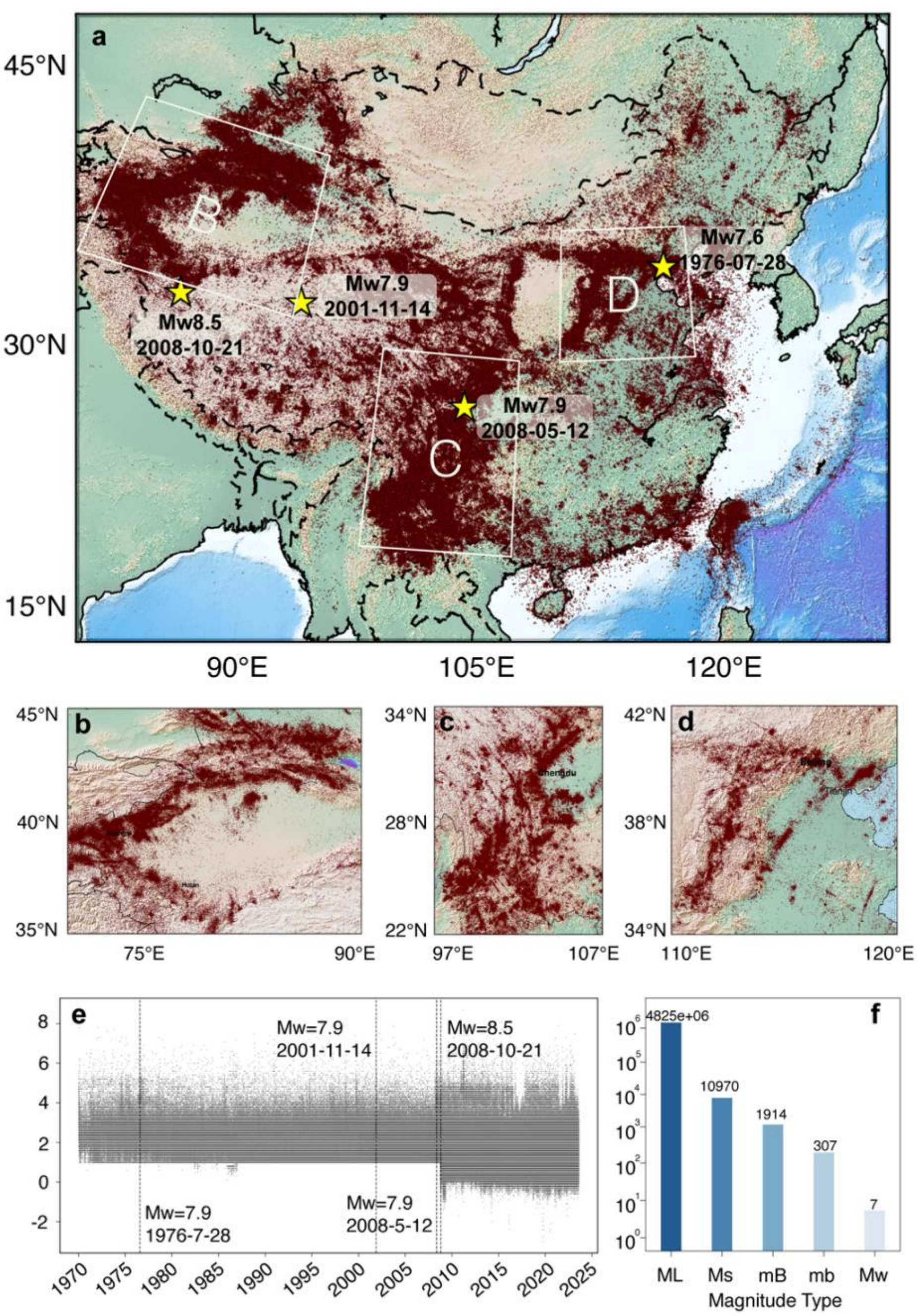

**Fig. 1** Distribution of earthquake events in mainland China from 1970 to 2024 and magnitude variation over time. (a) Distribution and zoning of seismic events in mainland China (dark red dots denote seismic events, yellow pentagrams indicate the largest seismic events within each zone, white trapezoidal boxes represent the three major zones); (b) Seismic distribution in the northwestern region; (c) Seismic distribution in the Sichuan-Yunnan region; (d) Seismic distribution in the North China Plain; (e) Distribution of seismic magnitude over time; (f) Frequency histogram of different magnitude categories.

## 2.1 Data Preprocessing

Catalog homogeneity is essential for reliable statistical seismicity analysis (Mousavi,

2017a). Since the CENC catalog records magnitudes across multiple scales (Fig. 1f), directly pooling these data would distort the FMD and bias the estimation of Mc and b-values (Bent, 2011). To ensure consistency, we converted all magnitude types into the physically well-defined, non-saturating moment magnitude Mw scale using established empirical relationships (Bormann & Saul, 2008; Cheng et al., 2017; R. Liu et al., 2007).

Furthermore, considering the significant spatio-temporal heterogeneities inherent in a unified catalog, a bulk analysis may obscure localized seismic characteristics. We partitioned the study area into key sub-regions—including the Northwest China, the North China Plain, and the Sichuan–Yunnan region—based on their distinct tectonic features (Fig. 1a, b, c, d).

Each region is analyzed on two complementary scales. For long–term analysis, we employ windows of 10–24 years to characterize stable seismic backgrounds. The partitioning of these windows is informed by the three distinct phases of CENC catalog quality identified by Mignan et al. (2013) — 1970–2001, 2001–2008, and 2008–present — a framework that mitigates the influence of network-generation effects on the inferred tectonic signal. We adopt this partition as a guiding reference rather than a rigid constraint, because strict adherence is not optimal in all regions. Where the Mignan phases would introduce additional catalog heterogeneity or obscure features of interest, we define windows that preserve a relatively uniform event density, as assessed from the magnitude–time distributions provided in the Supplementary Information. For the short-term analysis, windows of 1–5 years are centered on individual moderate-to-large events to resolve pre- and post-seismic evolution; the precise window bounds are specified in the caption of each corresponding figure.

### 2.2 Estimation of Mc and b-values

The accuracy of b-value estimation is highly sensitive to the estimation of the Mc. This study utilizes the Maximum Curvature (MAXC) method (Wiemer & Wyss, 2000) for its computational efficiency in localized time-window analysis. This nonparametric method requires fewer events than other techniques to achieve a stable result. However, the method underestimates Mc (Mignan & Woessner, 2012), and consequently, also the b value. To address this, an arbitrary correction factor of 0.2 is often added to the initial value; we term it $(Mc = M_{Maxc} + 0.2)$ (Woessner & Wiemer, 2005), resulting in an operational Mc of approximately 1.9 for most of our study areas. This value is highly consistent with the predicted future performance of the planned seismic network in the Chinese mainland as analyzed by Li et al. (2023). According to their findings, approximately 67% of the Chinese mainland currently maintains a posterior Mc < 2.0. Mc often exhibits strong spatiotemporal homogeneity (Mousavi, 2017b). Using a single, catalog-wide average can be heavily influenced by specific sub-regions with lower monitoring capabilities, leading to artificial inflation or deflation of b-values. Therefore, Mc is recomputed locally in every temporal window and every spatial neighborhood.

For b-value estimation, we employ the modified maximum-likelihood method (Aki, 1965; Utsu, 1965). All calculations for b-classical and b-positive indicators, including their temporal evolution, were performed using the SeismoStats package and were calculated on the same Mc thresholds. Temporal b-value curves are computed on a sliding event-window basis rather than a fixed-time sliding window. Each window contains N events and is advanced by one event at a time. Furthermore, we extended SeismoStats' functionality by developing a spatial mapping module. This addition enables high-resolution spatial mapping with a grid interval of 0.01◦ in latitude and longitude, as well as the visualization of b-values, a feature previously absent from the original toolkit. At each grid node, all events within a circular neighborhood of radius R = 0.5 ◦ (50km) are used.

We fully recognize that, for a statistical analysis paper such as ours, transparency in parameter choices is essential, as it enables the community to evaluate the reliability of our findings independently. The key parameters underlying our analysis are the time-window length (justified in the Data Preprocessing section), the minimum event count per estimate (N), and the spatial neighborhood criterion (a maximum search radius of 50 km), with the per-window N values listed in Table S1 of the Supplementary Information. Of these, N and the spatial-sampling threshold are treated as adaptive parameters because they exert a decisive influence on the reliability of the resulting estimates. Therefore, the new Section 4.1 documents a dedicated sensitivity test for these parameters, together with the statistical significance evaluation of the resulting b-value variations.

## 3 Results

### 3.1 Spatio-temporal Characteristics of b-values in Sichuan–Yunnan Region

Located within the lateral extrusion zone of the India–Eurasia collision, the Sichuan–Yunnan region is characterized by intense tectonic deformation and a complex network of active faults. It is one of the most seismically active areas in mainland China (Wen et al., 2008; Zheng et al., 2024).

Given the high complexity of the fault systems—particularly at major tectonic junctions where stress tends to concentrate—analyzing b-value heterogeneity is critical for seismic hazard assessment. To resolve these fine-scale spatio-temporal variations, we focus on four representative tectonic sub-regions (**Fig. 2**).

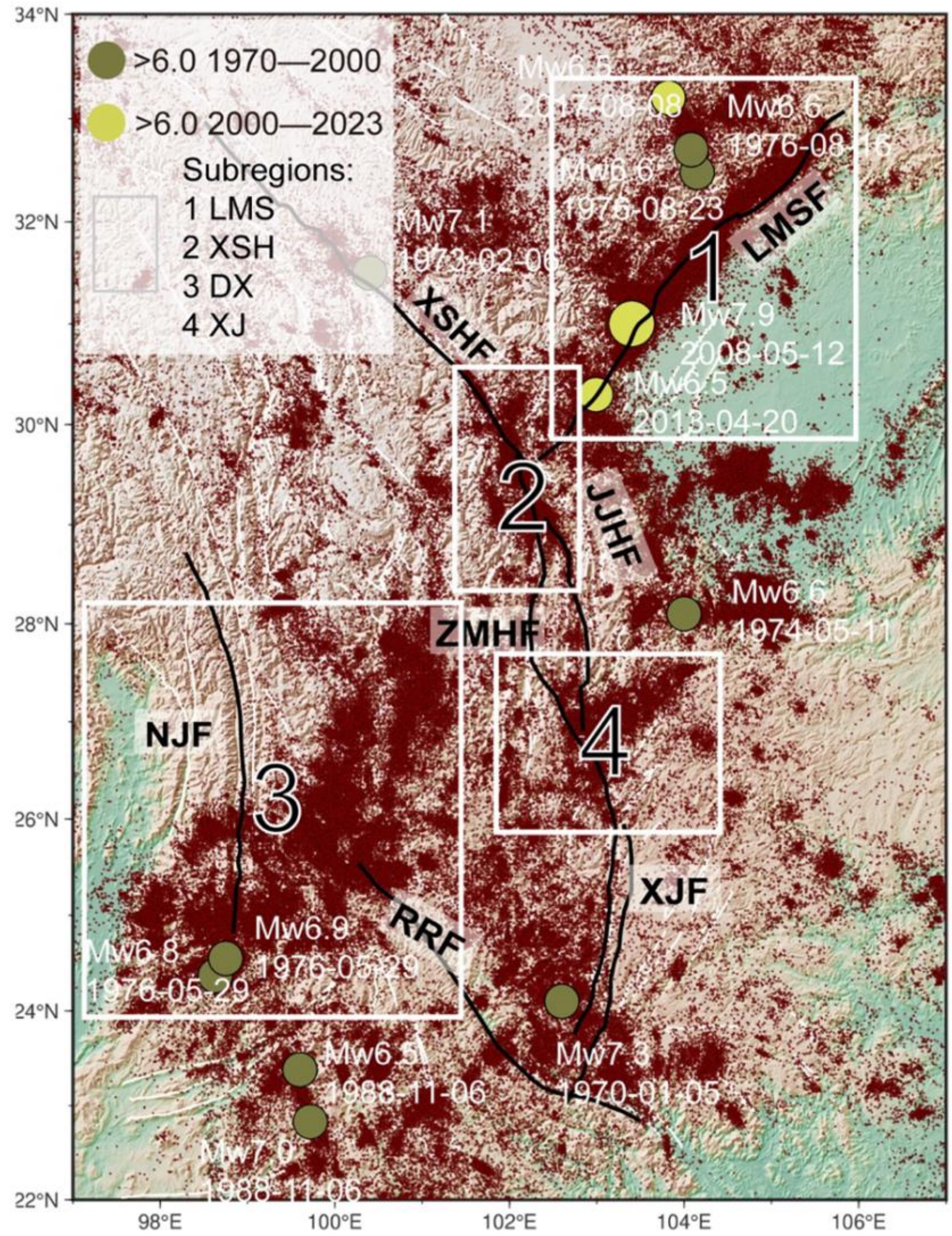

**Fig. 2** Distribution of seismic events and sub-regions in the Sichuan-Yunnan area. Dark green circles denote earthquakes of magnitude 6 or greater occurring before 2000; light green circles denote earthquakes of magnitude 6 or greater occurring after 2000; white dashed lines indicate primary and secondary fault zones; white rectangles denote the four delineated sub-regions. XSHF: Xianshuihe Fault; LMSF: Longmenshan Fault; JJHF: Jiaojihe Fault; XJF: Xiaojiang Fault; RRF: Red River Fault; NJF: Nujiang Fault.

### 3.1.1 The Longmenshan (LMS) Region

The Longmenshan (LMS) fault zone, situated at the eastern margin of the Tibetan Plateau, represents a prominent tectonic boundary between the Bayan Har Block and the Sichuan Basin (**Fig. 3**). In stark contrast to the rapid slip rates of the neighboring Xianshuihe–Xiaojiang fault, the LMS fault zone exhibits a conspicuously low strain rate. Geological investigations report a long-term average vertical slip rate of approximately 1.3–1.5 mm/yr (P.-Z. Zhang, 2013; Zheng et al., 2024), consistent with GNSS observations showing a cross-fault shortening rate of <2–3 mm/yr (Zhengkang et al., 2005). Despite this low loading rate, the LMS fault zone is capable of generating devastating great earthquakes, notably the 2008 Ms 8.0 Wenchuan earthquake (Deng et al., 2010; Liang et al., 2025) and the 2013 Ms 7.0 Lushan earthquake (Lei et al., 2014; Z. Liu et al., 2018).

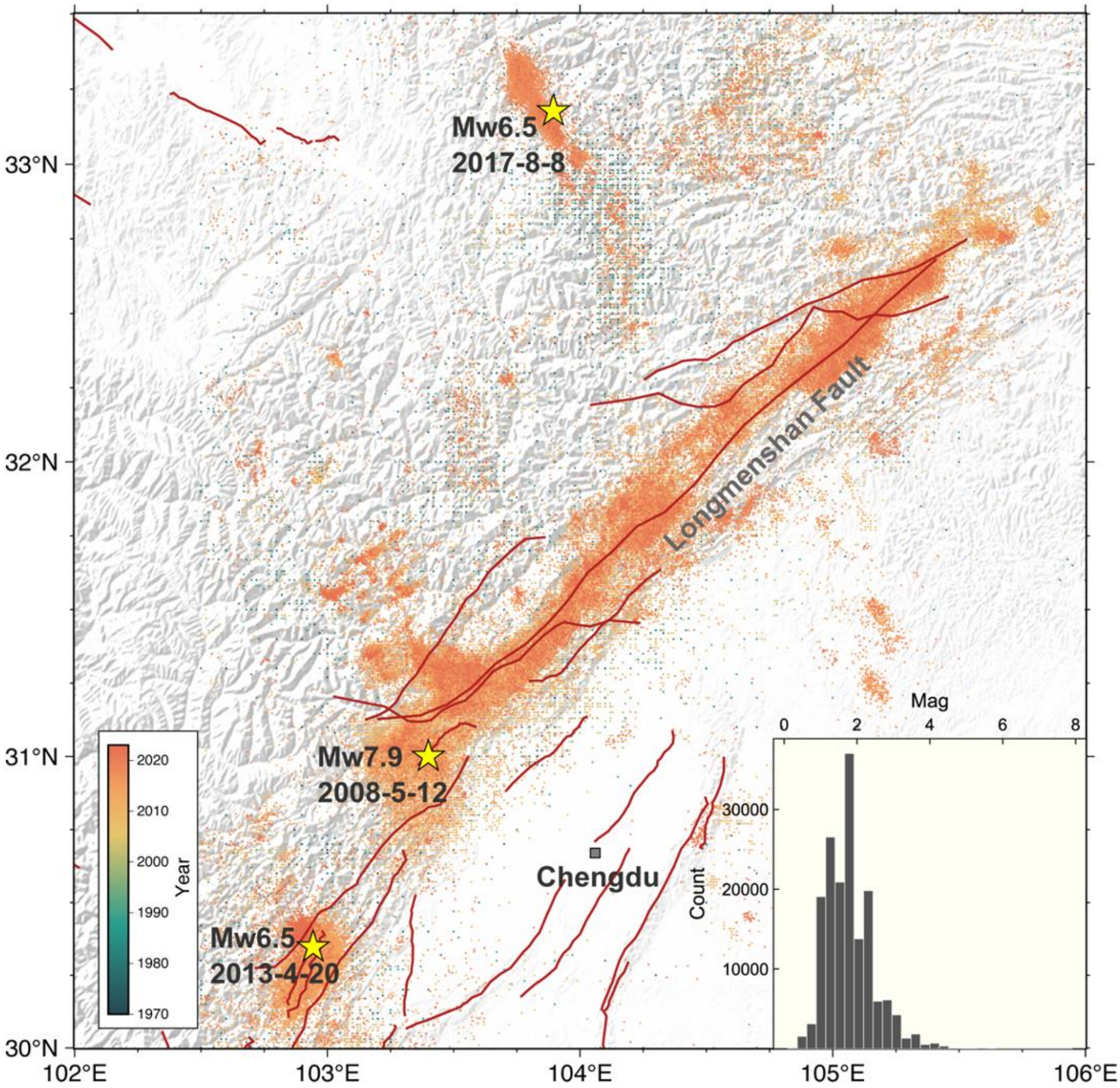

**Fig. 3** Distribution of seismic events and mainshock locations in the Longmenshan region, with a magnitude-frequency diagram. Earthquake times are color-coded by onset time; asterisks denote the three largest earthquakes by magnitude.

The magnitude–time (M-T) distribution for the LMS region from 1970 to 2023 (Fig. S1) clearly illustrates a phased evolution in seismic monitoring capability. To eliminate systematic biases arising from the varying detection thresholds, we partitioned the catalog into three distinct time intervals for independent calculation and analysis.

- **2006–2010: Spatio-temporal Evolution Associated with the Wenchuan Earthquake**

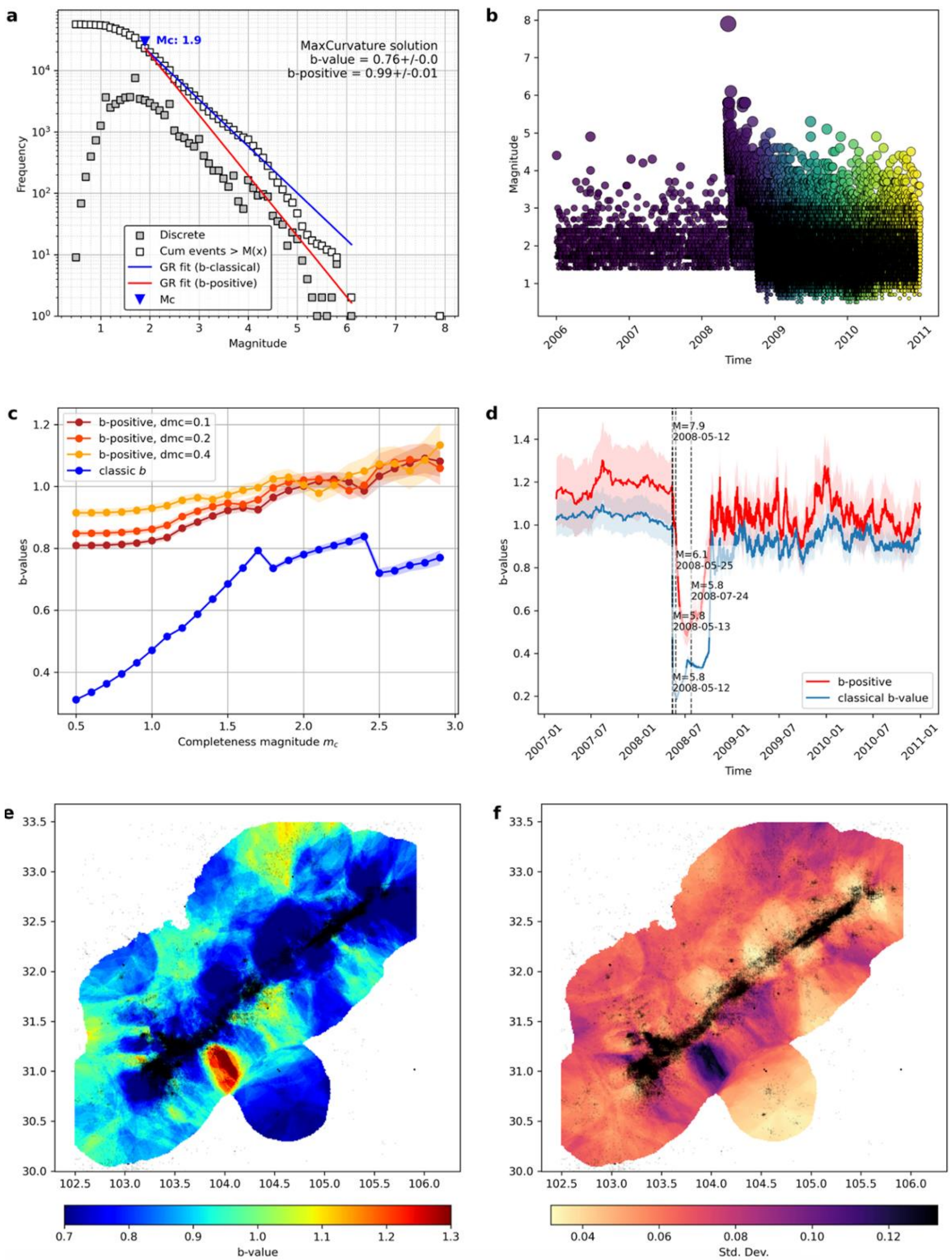

**Fig. 4** Spatio-temporal evolution of b-values in the Longmenshan (LMS) region from 2006 to 2010. (a) Frequency–magnitude distribution (FMD). (b) Magnitude–time distribution. (c) b-values as a function of Mc. (d) Temporal evolution of the b-values. (e) 2D spatial mapping of the b-classical. (f) Standard deviation (Std) of the 2D b-classical mapping.

Analysis of the short-period catalog covering the 2008 Wenchuan earthquake sequence yields a Mc of approximately 1.9 and a bulk b-classical of 0.76 for the LMS region. This b-classical is significantly lower than the global average (b ≈1.0), suggesting that the crustal medium was in a state of high stress accumulation or tectonic locking, which statistically favors a higher proportion of large-magnitude events. Before May 2008, b-values remained relatively high with minor fluctuations (b-positive ≈1.2; b-classical ≈1.0).

Zhao and Wu (2008) analyzed the 31-, 15-, and 10-year periods prior to the Wenchuan earthquake and found only a slight preseismic decrease near the epicenter. However, our results indicate that a catastrophic drop in b-values becomes evident when narrowing the observation to a short-term window leading up to the mainshock on May 12, 2008 (Fig. 4d). The reason contributing to the difference might be that long-term windows (such as the 10-year scale used by Zhao and Wu (2008)) may only capture subtle trends, not a catastrophic drop.

In our results, the b-classical plummeted to an exceptionally low range of 0.2–0.4, while b-positive decreased to approximately 0.5. Following the event, the b-values showed a rapid recovery, returning to a level near 1.0 by mid-to-late 2009. This trend indicates that the fault's stress state gradually re-equilibrated after the coseismic "shock."

The M-T distribution (Fig. 4b) vividly captures the aftershock decay following the 2008 Wenchuan earthquake. Notably, the volume of recorded microseismic events increased significantly over time, indicating pronounced temporal heterogeneity in the catalog. This shift in detection threshold raises the possibility that the observed b-classical trends are influenced by manufactured change (Mousavi, 2017a)—specifically, the marked enhancement of seismic network sensitivity. Therefore, the "catastrophic drop" in b-classical before the Wenchuan event warrants rigorous cross-validation. Encouragingly, the b-positive indicator, designed to correct this, strengthens the physical argument for stress accumulation.

However, the nuances in consistency between these two estimators will be further scrutinized in the following sections.

Spatial scanning results reveal a prominent linear low-b anomaly striking NE–SW along the Longmenshan fault zone, consistent with zones of high stress concentration (**Fig. 4**e). Such low b-values are indicative of high differential stress within strongly locked asperities, aligning with the "asperity mapping" technique established by Wiemer and Wyss (1997) who mapped low b-values (b≈0.5) in locked asperities (high stress) along the San Andreas Fault, versus high b-values in creeping sections. While creeping segments or fluid-rich zones typically exhibit high b-values due to distributed micro-seismicity, the low-b anomalies observed here explicitly map the "locked" portions of the LMS thrust system. It suggests that even at low geodetic strain rates, the high degree of coupling allows significant elastic strain energy to accumulate, ultimately facilitating cascading ruptures across multiple primary fault segments, as seen in the ~290-km-long 2008 Wenchuan event (Wang et al., 2001).

- **2011–2016: Post-Wenchuan Recovery and the Nucleation of the Lushan Earthquake**

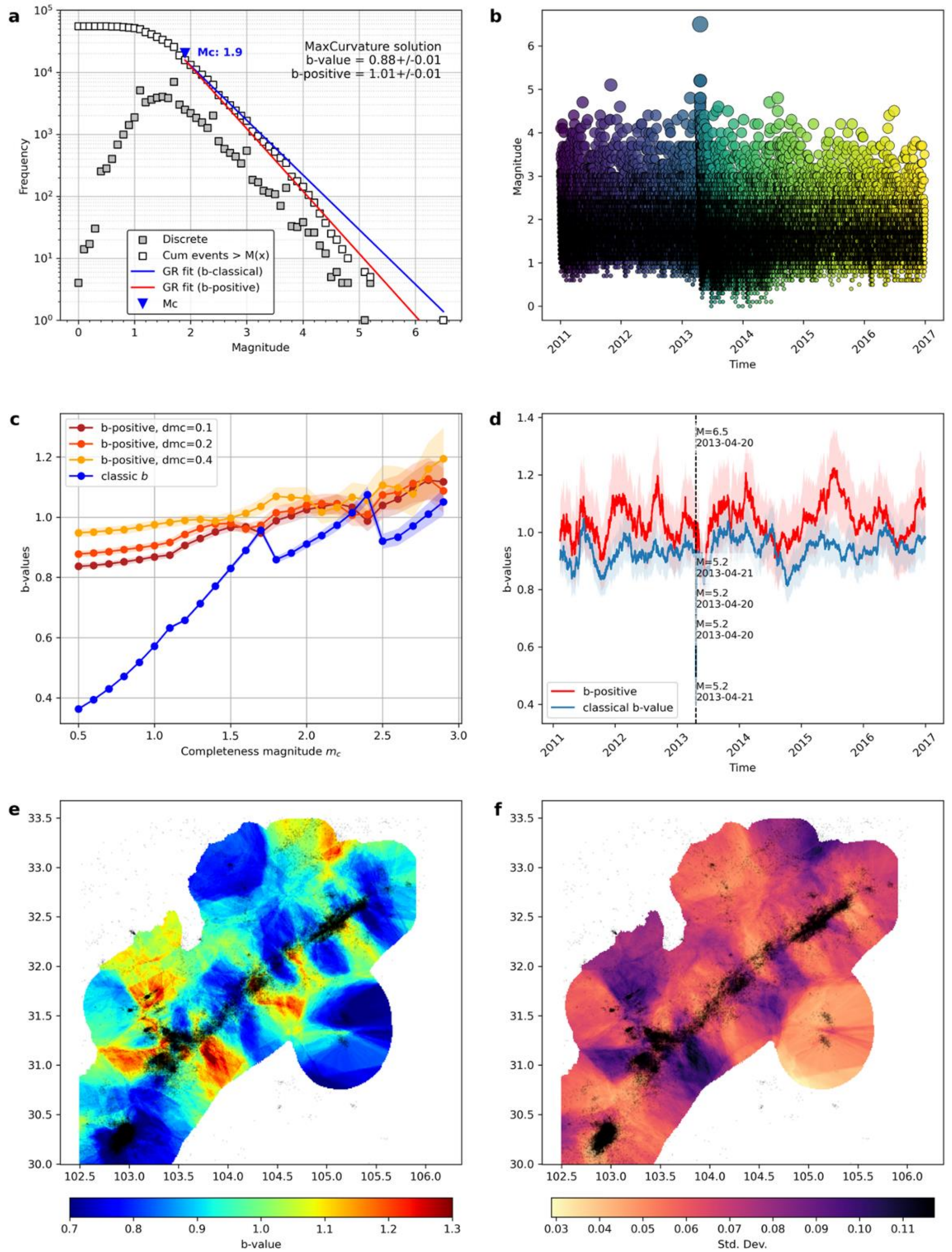

**Fig. 5** Spatio-temporal evolution of b-values in the Longmenshan (LMS) region from 2011 to 2016.

The 2011–2016 period represents the post-seismic recovery phase following the 2008 Wenchuan earthquake (Liang et al., 2025) and encompasses the April 20, 2013 Mw 6.5 Lushan earthquake (Lei et al., 2014). Statistical analysis yields a Mc of 1.9 for this interval. The classical G-R fitting gives a b-classical of 0.88, whereas b-positive is 1.01. This discrepancy suggests that, even years after a major event, varying computational weights assigned to small-magnitude events—driven by catalog incompleteness or internal fault-zone complexity—can lead to distinct estimator behaviors.

The temporal evolution of b-values primarily centers on the 2013 Lushan earthquake. We observe a transient perturbation in the b-classical curve around the mainshock. However, unlike the catastrophic drop preceding the Wenchuan earthquake, the decline here is less pronounced and exhibits a notable time lag. This "delayed" response may be attributed to the unusual occurrence of four M >5.0 aftershocks on the day of the mainshock. In contrast, before the November 1, 2011, Mw 5.1 event, both b-classical and b-positive show a synchronized decrease. Notably, during the July 2014 Mw 4.8 event, b-classical delayed while b-positive exhibited a more precise temporal response.

Spatial mapping reveals pronounced tectonic segmentation along the LMS fault zone during this period. The southern terminus (the Lushan source area) is characterized by a significant low-b anomaly, accurately delineating a high-stress asperity that facilitated the nucleation of the Lushan earthquake. Conversely, the central and northern segments—the primary rupture zones of the Wenchuan event—exhibit relatively high b-values, indicating that the crustal media remain fragmented and the stress release was more exhaustive. However, localized low-b patches embedded within this high-value background suggest an inhomogeneous fault healing process.

In summary, the 2011–2016 evolution record captures a dual tectonic process: the continued healing and aftershock decay in the central and northern segments, alongside the full nucleation-to-rupture cycle of the Lushan earthquake in the south. Of particular interest is a low-b anomaly ($< 0.8$) located toward the northern segment, which spatially correlates with the epicentral region of the August 8, 2017, Mw 6.5 Jiuzhaigou earthquake (Lu et al., 2022) and is discussed in the subsequent period (Figure S2).

3.1.2 The Xianshuihe (XSH) Region

The Xianshuihe fault zone, characterized by active seismicity and well-documented geological and seismic records (Allen et al., 1991), serves as an ideal laboratory for investigating seismic hazards (Fig. S2). This fault zone has undergone a protracted and complex process of tectonic evolution and deformation. Geological evidence indicates that over the past 10 million years, the Xianshuihe fault zone has experienced 90–100 km of sinistral (left-lateral) strike-slip displacement, with an average slip rate of (9.3±1.8) mm/year (Zhang, 2013; Zheng et al., 2024).

The magnitude-time (M-T) distribution of the Xianshuihe fault zone from 1970 to 2023 (Fig. S3) clearly exhibits temporal heterogeneity. Consequently, to effectively capture evolutionary characteristics ranging from the long-term background field to the short-term preseismic phase, this study selects two independent periods—2008–2023 (long-term window) and 2022–2023 (short-term window)—for comparative analysis.

The long-term window (2008–2023) provides a comprehensive record of the evolutionary history of the Xianshuihe fault zone over the past 24 years (Fig. S4, encompassing two significant tectonic events: the 2014 Kangding earthquake sequence (Jiang et al., 2015) and the 2022 Luding earthquake (Li et al., 2022; Zhao et al., 2023). Statistical analysis indicates that the Mc for this period is 1.9, with an overall calculated b-classical of 1.04.

Before the mainshock of the Mw6.1 Kangding earthquake sequence, the b-values (particularly the b-classical) exhibited a gradual decline over the years. This characteristic low b-values anomaly recovered rapidly following the earthquake, consistent with the classical "stress accumulation–release" cycle model. A similar long-term precursory decline was observed before the Luding earthquake.

- **2022–2023: Evolution of b-values in the Xianshuihe Region within the Short-term Window**

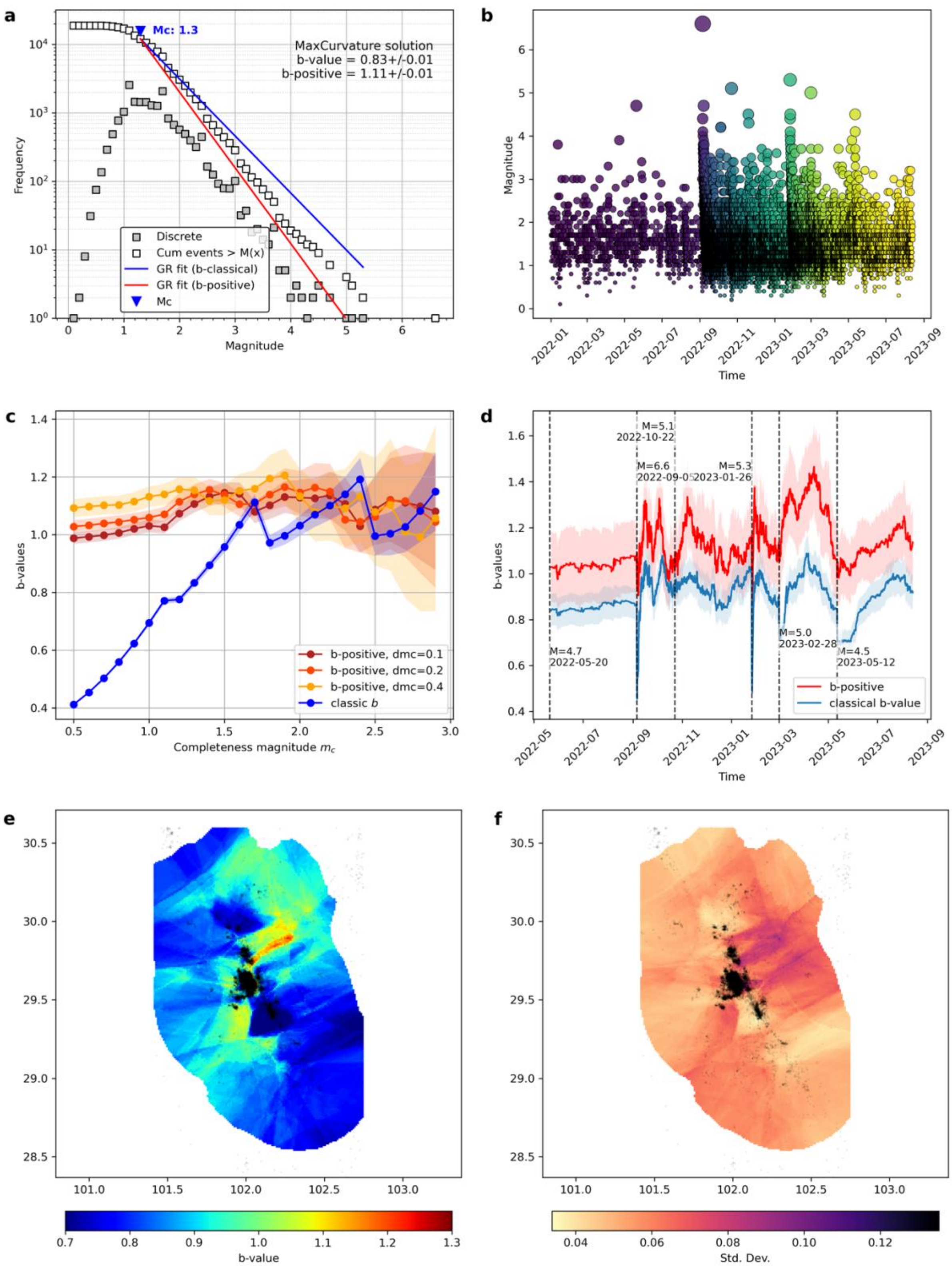

**Fig. 6** The spatiotemporal evolution of b-values in the Xianshuihe region from 2022 to 2023.

Within this short-term window, the b-classical is notably low at 0.83, whereas the b-positive value remains higher at 1.11. This significant discrepancy (0.83 vs. 1.11) is critical. It suggests that in the immediate aftermath of the Luding mainshock, the dense burst of numerous aftershocks led to waveform aliasing, resulting in severe short-term incompleteness of the earthquake catalog. It is clearly demonstrated in the M-T plot, where seismic records appear much sparser before the Luding event. While the classical method underestimates the b-value due to the missing records of small earthquakes, the b-positive method—utilizing higher-order magnitude differences—is inherently less sensitive to such data gaps. Consequently, the value of 1.11 is more representative of the actual physical state.

Most importantly, compared to the Longmenshan region, the b-values in the Xianshuihe region are markedly higher. Farrell et al. (2009) interpreted high b-value (b>1.3) variations as being related to stress changes accompanying the migration of magmatic and hydrothermal fluids. Similarly, Mousavi et al. (2017) attributed higher b-values to increased pore-pressure fluctuations, which facilitate micro-seismicity by reducing effective normal stress. Consequently, these elevated b-values may be attributed to the highly fractured or fluid-rich environment within the Xianshuihe fault zone, further strengthening the argument that the Xianshuihe region is mechanically distinct compared to the rigid North China block. Furthermore, the stress distribution patterns of strike-slip fault systems are likely to differ from those of thrust-type faults (such as the Longmenshan). Nevertheless, the b-values calculated for 2022–2023 are lower than those for 2000–2023, indicating that b-values in the Xianshuihe region have dropped below 1.0 in recent years.

3.1.3 Western Yunnan Region

- **2020–2023：Spatiotemporal evolution of b-values of the Yangbi earthquake sequence and evidence of fluid interaction**

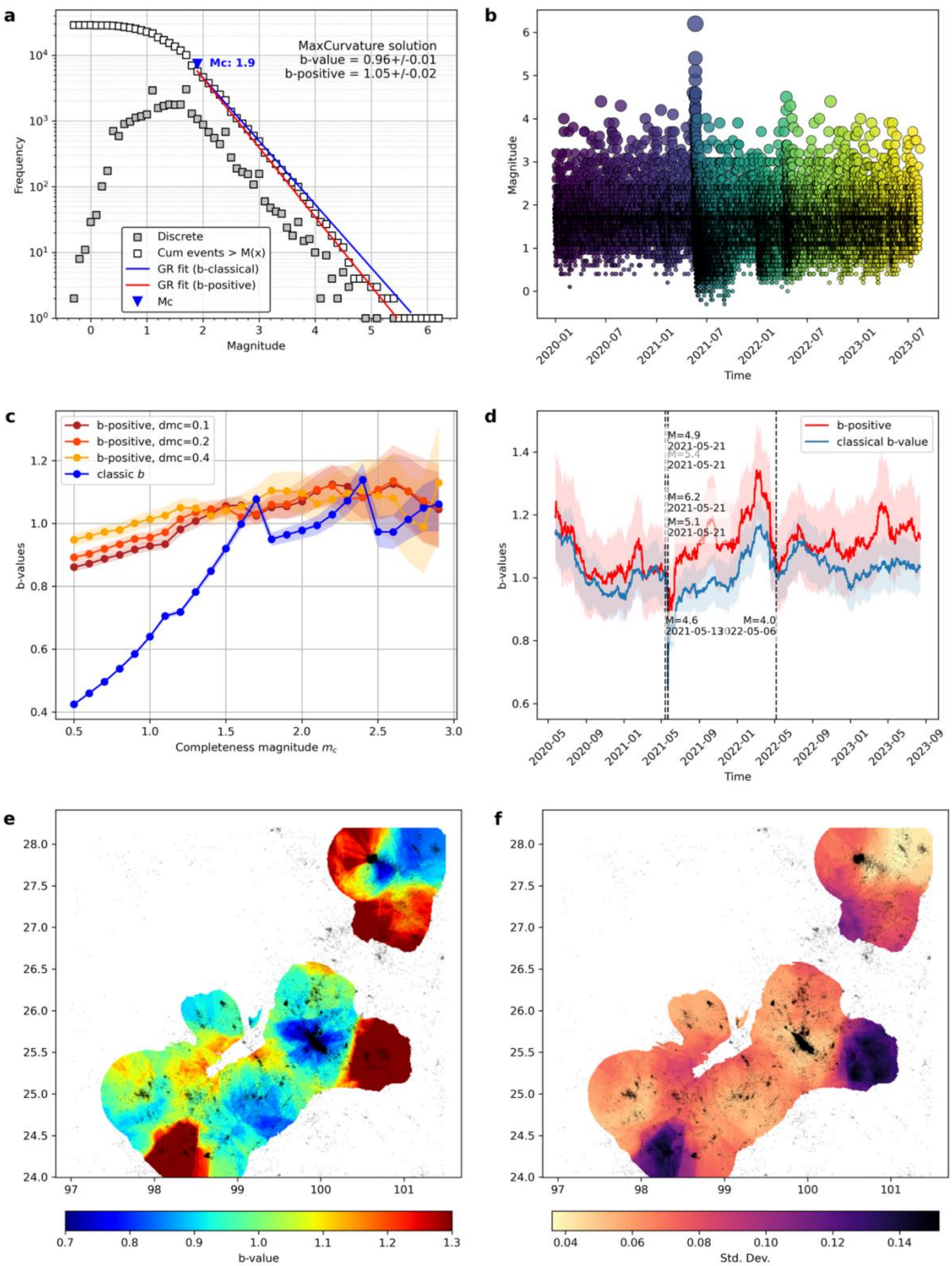

**Fig. 7** Spatio-temporal distribution of the b-values in the western Yunnan region from 2020 to 2023.

This time window focuses on the Mw 6.2 Yangbi earthquake (25.7°N,99.88°E) that occurred on May 21, 2021 (Duan et al. 2021) (Fig.S5) along with its foreshock-mainshock-aftershock sequence (**Fig.** ).

Benefiting from the coverage of a modern, high-density seismic monitoring network implemented after 2020, the calculated b-classical of 0.96 and the b-positive value of 1.05 exhibit a high degree of consistency during this period.

In the spatial scanning map, a highly distinctive source area structure is observed. The Yangbi epicentral area exhibits a deep-blue low-b-value anomaly that precisely delineates the main rupture zone or high-stress concentration area (asperity) of the Yangbi earthquake. More strikingly, the low-value core is closely surrounded by a prominent red-yellow "halo" of high b-values. This ring-like, high-b-value feature is typically interpreted as an aftershock diffusion zone with a high degree of fragmentation or as an area characterized by active, high-pore-pressure fluids. Given that existing studies suggest that the preparation of the Yangbi earthquake was closely related to fluid migration along the fault zone (Sun et al., 2022), this high b-value halo provides robust support for the fluid-involvement hypothesis.

### 3.1.4 Xiaojiang (XJ) Region

Located at the southeastern margin of the Tibetan Plateau (Fig. S8), the Xiaojiang fault zone (XJFZ) serves as the eastern boundary of the Sichuan-Yunnan rhombic block. It acts as a critical sinistral strike-slip structural belt regulating the southeastward extrusion of plateau material. Neotectonic activity in this region is exceptionally intense, characterized by high-frequency and high-magnitude seismicity. Historically, it hosted the catastrophic 1833 Songming Mw 8.0 earthquake (Shen et al., 2003) and more recent destructive events, such as the 2014 Ludian Mw 6.3 earthquake (Riaz et al., 2017).

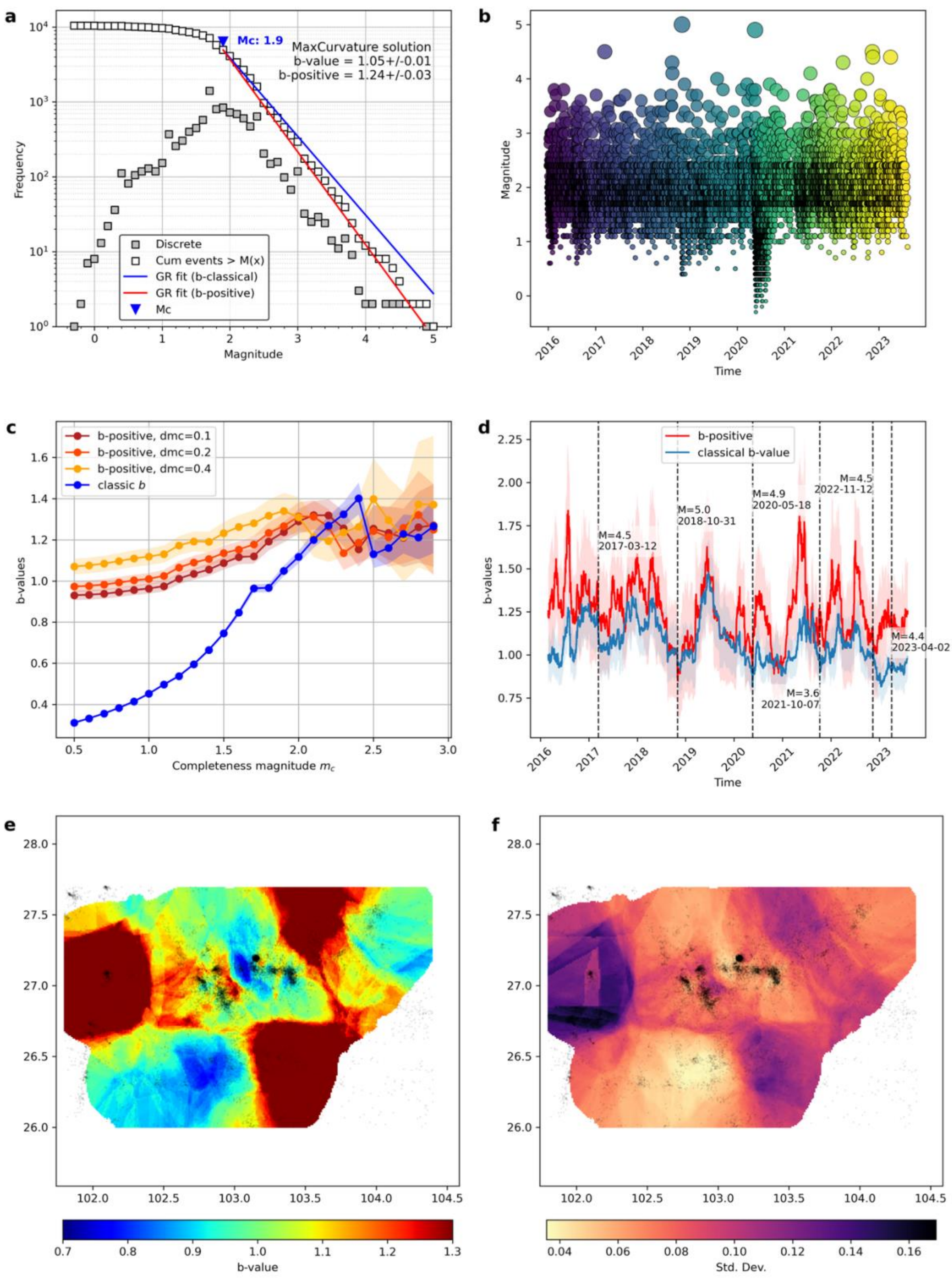

**Fig. 8** Spatiotemporal evolution of b-values in the Xiaojiang region from 2016 to 2023.

From a short-term perspective, spanning the 7-year window from 2016 to 2023, the b-classical has consistently remained within a low-value plateau between 0.8 and 0.9, exhibiting minimal fluctuation. It represents a typical signal of strong fault locking, indicating exceptionally high normal stress on the fault plane. Under such conditions, microfracturing is suppressed, and energy accumulates rapidly.

The spatial distribution map likewise reveals a fragmented, patch-like pattern. A prominent deep-blue low b-value anomaly is present in the central portion of the map, corresponding to the locked segments of the Xiaojiang fault zone or the Qiaojia-Dongchuan seismic gap. Compared to the linear characteristics observed in the Longmenshan fault, the alternating high and low b-values in the Xiaojiang region are more complex. It suggests the interaction of multiple secondary faults and implies that stress concentration points are relatively dispersed.

In summary, the Xiaojiang fault zone exhibits a non-homogeneous mosaic pattern spatially characterized by "localized strong locking (low b-value cores) encased within a broad background of fractured medium (high b-values)." This tectonic configuration suggests that although the 2014 Ludian earthquake may have partially released regional accumulated stress—thereby reducing the immediate probability of a catastrophic earthquake—the stress concentration phenomena dominated by localized locked units remain significant. Consequently, the risk of moderate-to-strong earthquakes (approximately magnitude 5.0) in this region should not be overlooked.

### 3.2 Spatio-temporal Characteristics of b-values in North China Plain

The North China Plain, situated in the eastern portion of the North China Craton (NCC), is a Cenozoic rift basin developed upon an ancient metamorphic basement (Fig. S10). The far-field geodynamics of the westward subduction of the Pacific Plate profoundly influence this region. Its geological structure is primarily characterized by extensional rifting and block subsidence, with a series of developed north-northeast (NNE) trending blind faults. Unlike the high-frequency plate-boundary activity observed in the Sichuan-Yunnan region, the North China Plain exhibits the typical characteristics of a "rigid block" within a continental intraplate setting. Despite a relatively low level of background micro-seismicity, it possesses a high capacity for stress accumulation. Historically, this region has hosted catastrophic events such as the 1679 Sanhe-Pinggu magnitude 8.0 earthquake (Ran et al., 1997) and the 1976 Tangshan magnitude 7.8 earthquake (Butler et al., 1979), demonstrating a tectonic background capable of generating high-stress-drop, devastating earthquakes. Consequently, it serves as a core area for seismic hazard monitoring within the Capital Metropolitan Region.

From a long-term perspective, the b-classical is only 0.64, whereas the b-positive value reaches 1.02 (Fig. S11). The massive chasm between these two figures (0.64 vs. 1.02) illustrates severe non-homogeneity in the regional earthquake catalog over a 50-year timescale. In the early periods, sparse monitoring stations led to the omission of numerous small earthquakes, artificially depressing the classical b-value. However, even after correction, the overall b-value (approximately 1.0) is not considered high for an extensional zone, suggesting a high level of background stress.

This uniform low b-value distribution is a hallmark of a rigid block. In contrast to the fragmented nature and frequent microseismicity of the Sichuan-Yunnan region, the crust of the North China Plain maintains high integrity and strength. It implies that the region typically experiences fewer small earthquakes, making energy dissipation difficult; once a rupture occurs, it tends to manifest as a high-magnitude, high-stress-drop destructive earthquake.

- **2008-2023 Evolution of b-values Characteristics in the North China Plain within the Short-term Window**

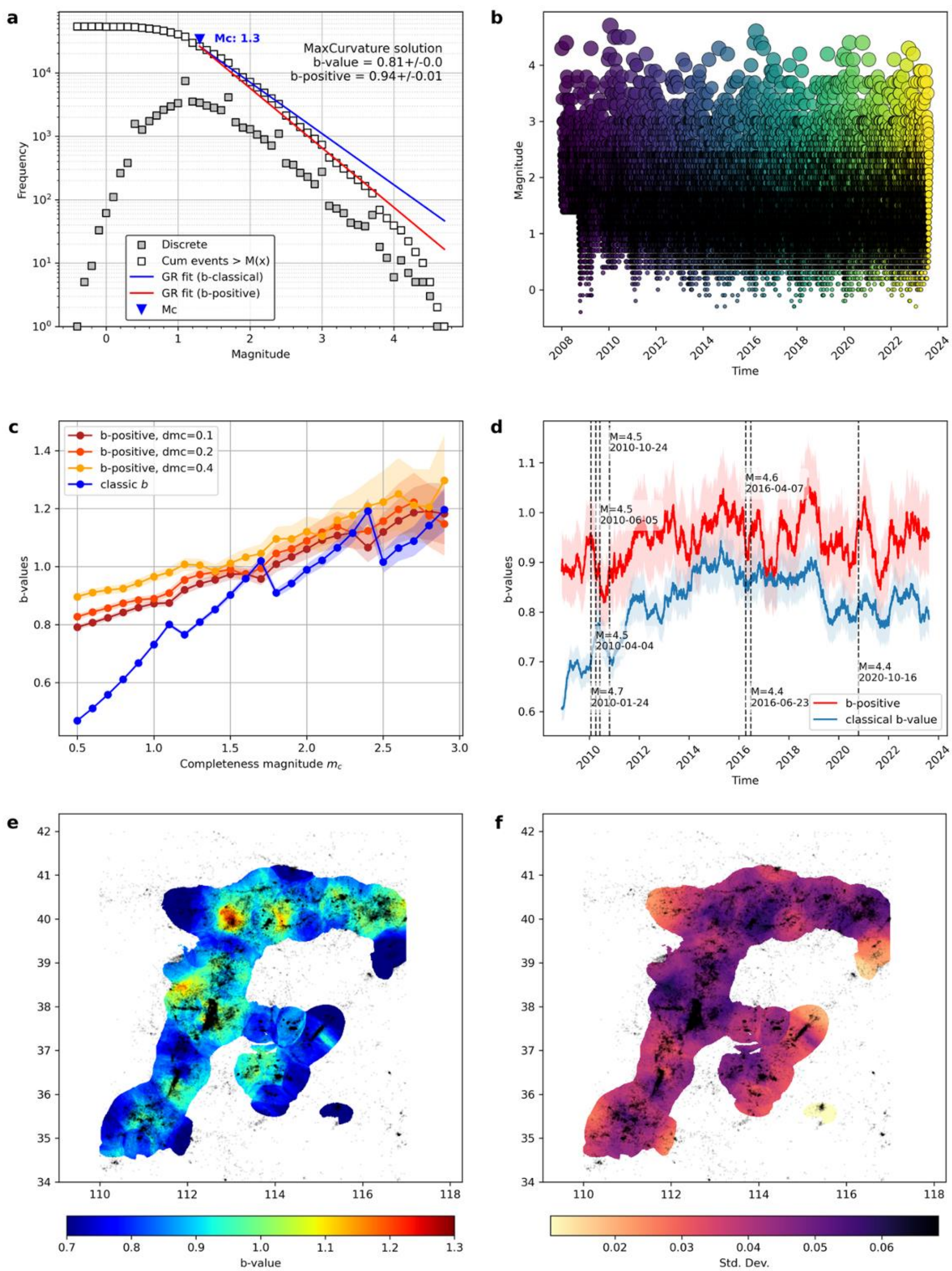

**Fig. 9** The spatiotemporal evolution of b-values in the North China Plain from 2008 to 2023.

When the time window is narrowed to the most recent two years and observed at a more microscopic scale, the deep-blue regions that previously appeared as uniform anomalous low-value zones (Fig. ) reveal a complex internal structure. It is partly attributed to the increased density of the monitoring network, which has reduced the Mc to 1.3.

The b-classical has recovered to 0.81, significantly narrowing the gap with the b-positive value of 0.94. While this indicates that previous extremely low values were partially derived from data gaps, the value of 0.81 remains lower than the Sichuan-Yunnan region (>1.0), confirming a higher background stress level. The heightened stress state is intrinsically linked to the tectonic framework of the NCC, which is characterized by a complex network of NNE-striking lithospheric faults and NW-striking buried faults (Kusky et al., 2016). Specifically, the North China Plain resides within the North China Basin, a region defined by extensive Cenozoic rifting and active faulting that segments the crust into distinct sub-blocks.

Despite the occurrence of only small-to-moderate earthquakes of magnitude 4.0 or less during this period (Fig. ), the b-value curves exhibit exceptionally high sensitivity to these events. Before nearly every marked magnitude 4 event, a "pulse" pattern—characterized by an initial rise followed by a rapid decline in b-values is observable. However, the magnitude of these variations is far smaller than the abrupt mutations observed in the moderate-to-strong earthquakes discussed previously.

In contrast to the uniform blue of the long-term period (Fig. S11), the short-term spatial map shows a complex texture of interleaved red and blue zones. The significantly lower overall b-values compared to those in the Sichuan-Yunnan region confirm that the crustal medium of the North China Craton possesses high strength and good homogeneity. Unlike the fragmented energy release observed at the margins of the Tibetan Plateau, the NCC's rigid block architecture facilitates high levels of stress accumulation. Such a tectonic environment is unfavorable for the release of energy in discrete events and maintains the background conditions required to generate large earthquakes. By identifying blue-locked patches in the short-term maps, it is possible to delineate specific hazardous zones for future earthquakes of magnitude 4 to 5, or even stronger.

### 3.3 Spatio-temporal Characteristics of b-values in the Northwest Region

The Northwest region encompasses the Tianshan-Kunlun seismic belt (including Xinjiang and its adjacent areas), a domain characterized by a typical Basin-and-Range structural configuration (Fig. S12). Seismic activity is primarily concentrated within the rigid orogenic belts—specifically the Tianshan, Kunlun, and Pamir Plateau—while the interior of the intervening Tarim Basin remains relatively stable.

- **Post-2006 Seismicity**

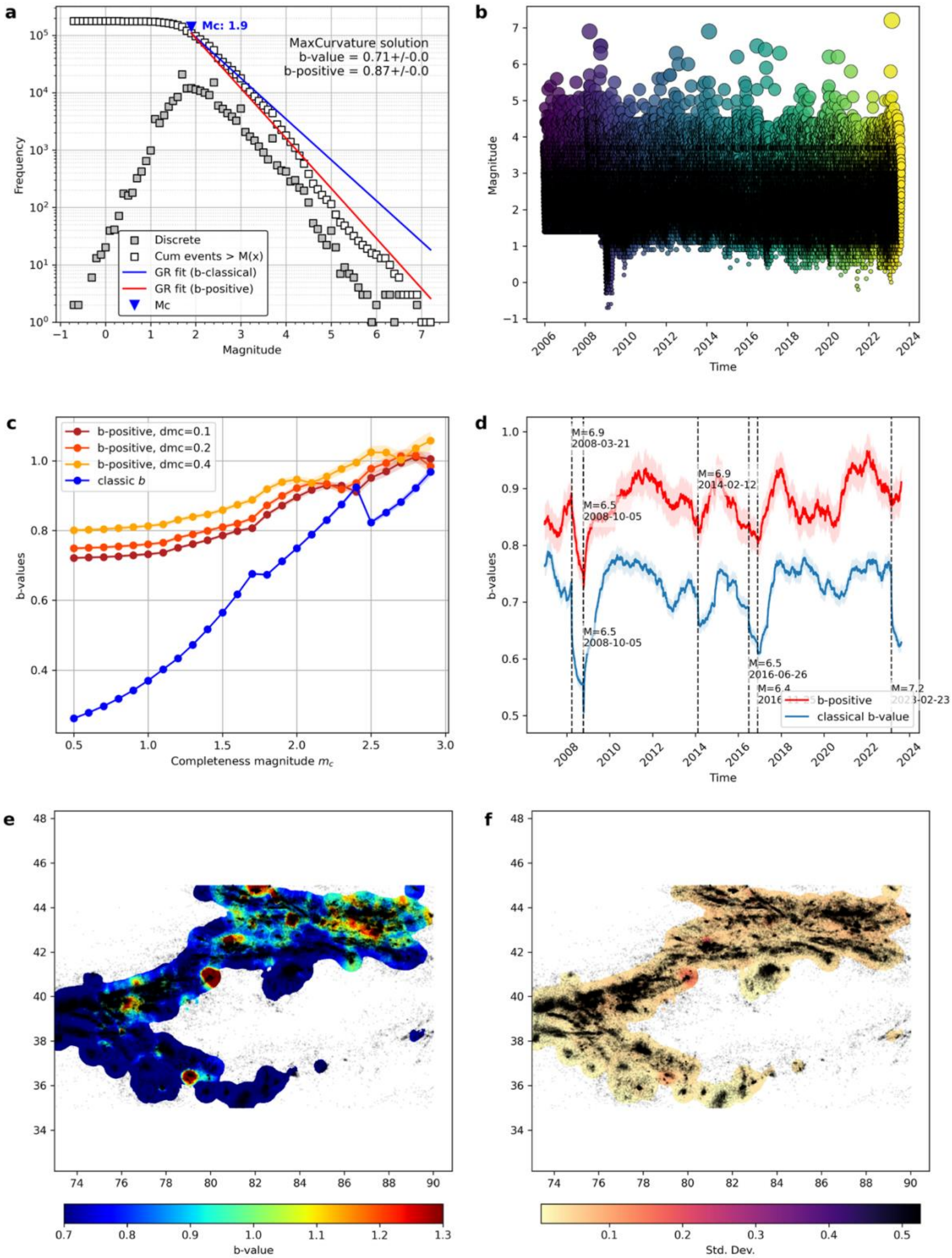

**Fig. 10** The spatiotemporal evolution of b-values in the Northwest Region from 2006 to 2023

Since 2006, this region has hosted several earthquakes with magnitudes exceeding Mw 6.5, identifying it as a high-stress, high-magnitude seismogenic zone. This period covers multiple major events of magnitude 7 or greater, including the 2008 Yutian, 2014 Yutian, and the 2023 Tajikistan-Xinjiang border earthquakes. Under the modern monitoring network, Mc has stabilized at 1.9. Despite this improved data completeness, the b-values remain markedly low, with the b-classical at 0.71 and the b-positive value at 0.87. It confirms that a low b-value is an inherent tectonic attribute of this region rather than a mere artifact of data error. Notably, the b-values curves exhibited distinct precursory sudden drops preceding each of these earthquakes of magnitude 6.5 or greater. The abrupt decline before the 2008 Mw 6.9 earthquake was particularly significant.

The spatial distribution map displays the deepest black-blue patches in the lower-left corner. This area corresponds to the intersection of the Pamir Plateau and the Western Kunlun Mountains. It represents one of the regions on the Chinese mainland currently experiencing the most intense compressive stress. In contrast, the Tianshan seismic belt to the north exhibits an interleaved blue-and-yellow pattern. It indicates varying levels of activity across different segments of the Tianshan fault zone, with some segments remaining locked (blue) while others are relatively fractured and active (yellow).

## 4 Discussions

### 4.1 Sensitivity Testing and Uncertainty Analysis

The reliability of the spatiotemporal patterns reported in Section 3 depends on the choice of the completeness magnitude Mc, the sliding-window size N, the spatial neighborhood, and the statistical framework used to evaluate temporal change.

Parameter choices for every window analyzed in the main text — Mc, N, the b-classical and b-positive estimates with their bootstrap standard deviations, and the corresponding b-significant p-values — are listed in Table S1, providing a single reference for the methodological inputs behind every b-value reported in this paper. Per-window FMDs, Mc-sensitivity curves, and per-cell bootstrap uncertainty maps are shown in panels (a), (c), and (f) of each regional figure. Across all regions, the b-positive estimator is markedly less sensitive to Mc than the classical estimator, consistent with the construction of b-positive from successive magnitude differences(van der Elst, 2021).

To demonstrate that the principal anomalies are not artifacts of any single parameter choice, we conducted a full multi-parameter sensitivity test on the Longmenshan 2006–2010 window as a worked example. The temporal b-value series was recomputed across three sliding-window sizes and three Mc values (Table S2, Figures S15); the 2D b-value map was recomputed across six combinations of grid spacing, search radius, minimum event count per grid node, and Mc (Table S3, Figure S16). The preseismic decrease of b-positive before the Wenchuan event and the low-b patch coinciding with its rupture area are recovered in every configuration. The protocol can be applied to any other window upon request.

Statistical significance of temporal variation is evaluated using the b-significant autocorrelation method of (Mirwald et al., 2024), which tests the null hypothesis of a constant b-value using the autocorrelation of a sliding-window series. A fluctuation is treated as a structured tectonic signal only if the null hypothesis($H_0$) of the constant b-values calculated from N over time is rejected at $p < 0.05$. This approach ensures that the reported drops or "pulses" in b-values are statistically significant rather than artifacts of small-sample noise. Detailed results for each window, including specific N values and p-values for b-classical and b-positive, are provided in Table S1 in the Supplementary Information. Specifically, the p-values were sufficient to reject the null hypothesis of a constant b-value over time, confirming that the observed fluctuations represent robust tectonic signals rather than random noise.

## 4.2 Distribution Characteristics of b-values in Major Tectonic Regions of China

The distribution of b-values across the Chinese mainland exhibits significant spatial heterogeneity, broadly reflecting the region's diverse tectonic regimes. Fault zones in the Sichuan-Yunnan region display diverse characteristics, ranging from low b-values along the Longmenshan fault to higher background values along segments of the Xianshuihe and Red River faults. These regional observations align with the global finding that b-values vary systematically by faulting style: normal faulting typically yields the highest b-values ($b \approx 1.1$), strike-slip faults exhibit intermediate values ($b \approx 0.9$), and thrust faults show the lowest ($b \approx 0.7$) (Schorlemmer et al., 2005). This global hierarchy provides a physical baseline for interpreting the mechanical states of the Sichuan-Yunnan fault systems.

In the Longmenshan region, low b-value anomalies ($b<0.8$) are distributed in clear, linear bands along the fault of strike, consistent with the low b-values expected for a strongly locked thrust-compression system, in agreement with the geodetic coupling estimates of Shen et al. (2005) and Zheng et al. (2024) for this segment. In contrast, the Xianshuihe region consistently exhibits a higher background b-value (approximately 1.04). This higher value not only reflects the inherent intermediate baseline of strike-slip regimes but also may suggest medium heterogeneity and possible fluid involvement. Despite the high overall b-value, localized low-b-value patches can be identified along the fault in both long- and short-term maps. In a retrospective sense, several of these patches spatially coincide with the rupture areas of subsequent moderate-to-large earthquakes; for instance, the 2022 Luding earthquake occurred within a localized low-value zone, where the b-value was significantly lower than the regional strike-slip average.

Furthermore, monitoring data from the past two years indicates that the b-value in this region has dropped significantly below 1.0. Rather than exhibiting a long-term decline before the 2022 Luding earthquake, it showed a statistically significant transient decrease ($p < 0.05$; Table S1) in the months immediately before the event. It suggests that the fault zone may currently be in a phase of rapid critical stress accumulation, consistent with a transition from creep to locking. However, this interpretation requires independent geodetic confirmation. Similarly, the Xiaojiang fault zone, located at the margin of the Sichuan-Yunnan rhombic block, presents a complex "mosaic" tectonic configuration. A 20-year low b-value plateau (0.8–0.9) emits a strong signal of tectonic locking. Its non-homogeneous spatial distribution—characterized by "localized strong locking (low b-value) encased within a broad background of fractured medium (high b-value)"—is consistent with the observation that, while the overall regional fragmentation is high, localized stress concentrations may persist along the southern segment, where secondary faults intersect.

In the Western Yunnan region, the 2021 Yangbi earthquake serves as a quintessential example, with the source area exhibiting a characteristic structure: a main rupture core defined by low b-values, surrounded by a peripheral "halo" of high b-values associated with fluid migration and aftershock diffusion. This observation provides independent statistical evidence for the previously proposed hypothesis that fluid-driven processes contribute to fault weakening (Sun et al., 2022). Furthermore, historical data comparisons in this region powerfully demonstrate the advantages of the b-positive method in addressing long-term catalog non-homogeneity, effectively correcting the artificially low classical b-values caused by insufficient monitoring capabilities in earlier periods.

Regarding intraplate tectonic zones, the North China Plain is characterized by a generally low b-value distribution; however, even the corrected b-value (approximately 1.02) does not exhibit the high-value characteristics typically expected in extensional environments (where b-values usually exceed 1.1). In the Northwest region (Tianshan-Kunlun seismic belt), the extremely low b-value (approximately 0.71) is consistent with an inherent tectonic attribute of the 'basin-range coupling' system, and the multi-parameter sensitivity test in Section 4.1 (Table S2-3, Figure S15-16) confirms that this low value is not an artifact of any single parameter choice. The Pamir-Western Kunlun tectonic knot exhibits the lowest b-values in the entire region, in a region of intense plate convergence and compressive action.

## 4.3 Reliability Analysis of b-value as a Large Earthquake Precursor

Several studies have argued that apparent temporal variations in b-value are often artifacts of network changes, undersampling, or analytical inconsistencies (Görgün, 2013; Kamer, 2014). Critiques such as those by Kamer and Hiemer (2013) emphasize that, without rigorous statistical significance testing, precursors may be indistinguishable from random fluctuations in the catalog, which is "a mere artifact of the sliding time window". Although the effectiveness of the b-value as an earthquake precursor remains a subject of long-standing debate and is neither a sufficient nor necessary condition for seismic occurrence, the present study provides further empirical evidence that, when computed using estimators robust to catalog incompleteness and evaluated under formal statistical-significance criteria, b-value variations carry information about the regional state of stress.

By addressing these concerns using the b-positive method, which is specifically designed to mitigate biases introduced by network heterogeneity, this study observes statistically significant decreases in b-values ($p < 0.05$ in the b-significant test; Table S1) preceding several moderate-to-large earthquakes across different regions. In addition to results consistent with previous studies for the Wenchuan earthquake (Liang et al., 2025), this paper provides further evidence from other regions and time windows. For instance, a significant sudden decline in b-values was observed before multiple major historical earthquakes (e.g., Wenchuan, Tangshan, and Ludian), consistent with the inverse stress-b relation proposed by Scholz (1968, 2015)and Goebel et al. (2013).

From the perspective of long-term spatiotemporal evolution, the Jiuzhaigou earthquake provides a highly compelling case study. Spatiotemporal scanning reveals that the northern segment of the Longmenshan fault zone (the Jiuzhaigou seismic area, 33.2° N, 103.8°E) exhibited a prominent deep-blue low b-value anomaly ($b < 0.8$) between 2011 and 2016 (**Fig. 5**e). This anomalous zone subsequently hosted the 2017 Jiuzhaigou Ms 7.0 earthquake, after which the values transitioned to a high-value state ($b > 1.1$) (Figure S6), consistent with the post-mainshock stress-drop signature reported by Gulia & Wiemer (2019). This sequence of observations — a prolonged low-b anomaly followed by a moderate-to-large earthquake within the anomaly and a post-seismic transition to high b — provides a retrospective, single-case illustration of the spatial-temporal pattern that motivates the use of b-value mapping in hazard assessment. We emphasize that a single retrospective case does not establish prospective forecasting skill; a formal pseudo-prospective evaluation against an independent test catalog would be required for that claim. In a retrospective sense, the Jiuzhaigou case illustrates how long-term b-value mapping can complement, but not replace, conventional geodetic and historical-seismicity-based hazard assessment.

## 4.4 Robustness Analysis of the b-positive Method

A comparative analysis between the classical Gutenberg–Richter MLE and the b-positive method shows that the b-positive estimator yields more stable b-value time series under the catalog conditions of mainland China. In several of the cases examined, b-positive resolves a temporal decrease prior to moderate-to-large earthquakes that the classical estimator either misses or misattributes to network changes. It demonstrates enhanced stability when processing incomplete data caused by changes in seismic network configurations or by bursts of short-term aftershocks (e.g., the red line in Fig. 3d and **Fig. 4**d), making it better suited for real-time or quasi-real-time seismic trend tracking.

The pre-1980 portion of the catalog provides a diagnostic illustration of how catalog incompleteness can produce spurious b-classical anomalies. In the Western Yunnan region (Fig. S7d), b-classical remained anomalously low (≈ 0.25) before the May 29, 1976, Mw 7.1 event and then rose abruptly to ≈ 0.75 around the time of the mainshock — the opposite of the precursor pattern that b-classical is sometimes proposed to display. The abrupt rise in 1980 (Fig. S5b) coincides with the documented improvement of the regional network, indicating that the apparent low b-classical reflects incompleteness rather than tectonic state. Similar pre-1980 anomalies in b-classical are present in our Longmenshan, Xianshuihe, North China Plain, and Northwest records (Fig. S1d, S3d, S11d, S13d), all of which coincide with the network-generation transitions identified by Mignan et al. (2013). We therefore restrict tectonic interpretation of b-classical to the post-2008 portion of the catalog, where network coverage is comparable across regions; b-positive shows correspondingly smaller pre-1980 fluctuations, but, given the very limited event counts in this period, we do not interpret either estimator's pre-1980 values as a tectonic signal.

Across the post-2008 cases analyzed in this study, b-positive remained stable under varying network conditions and resolved temporal decreases prior to several moderate-to-large earthquakes that passed the b-significant test ($p < 0.05$; Table S1), including the 2008 Wenchuan, 2017 Jiuzhaigou, and 2022 Luding events. In other cases — for example, the secondary fluctuations associated with $M \lesssim 4$ events in the North China Plain — neither estimator yields a statistically significant pre-event decrease, and these cases are described in the text as observations only. A formal pseudo-prospective evaluation against an independent test catalog would be required before b-positive could be recommended as an operational forecasting indicator.

**5 Conclusions**

This study presents the first systematic, mainland-wide retrospective evaluation of b-value variations in China as a tool for earthquake precursor monitoring, based on a homogenized 1970–2024 catalog re-projected onto a single magnitude scale (Mw). Six tectonically distinct sub-regions — the Longmenshan thrust system, the Xianshuihe strike-slip system, the western Yunnan and Xiaojiang fault zones, the North China Plain, and the Northwest Tianshan–Kunlun orogen — are analyzed under a unified framework, allowing for the first time a direct cross-regional comparison of b-value behavior across mainland China.

Our principal findings are: (i) the long-term spatial b-value distribution shows clear, tectonically meaningful contrasts between thrust, strike-slip, and rift environments, with low-b patches that, in a retrospective sense, spatially coincide with the rupture areas of several moderate-to-large earthquakes in the catalog, including the 2008 Wenchuan, 2017 Jiuzhaigou, and 2022 Luding events; (ii) localized high b-value anomalies in regions such as western Yunnan are consistent with regions previously interpreted as having fluid involvement, supporting the long-standing interpretation of high b as a signature of fluid involvement and medium heterogeneity; and (iii) statistically significant decreases in b-positive precede several of these events, while the corresponding decreases in b-classical cannot, in some cases, be cleanly separated from contemporaneous changes in network completeness.

A direct, head-to-head comparison of the classical maximum-likelihood estimator and the b-positive estimator under Chinese continental conditions reveals that b-positive is substantially less sensitive to Mc, less affected by short-term aftershock incompleteness, and more stable across long-term catalog-quality transitions. Our results support the b-positive estimator as the more robust diagnostic for precursor monitoring in catalogs with nonuniform completeness histories, while emphasizing that the present analysis is retrospective and not a prospective forecasting demonstration.

Additionally, this work provides a transparent, reproducible benchmark for b-value analysis of large continental catalogs. For every temporal window analyzed in the main text, Table S1 reports the local Mc, the sliding-window size N, the spatial-mapping minimum event count Map_N, the b-classical and b-positive estimates with their bootstrap standard deviations, and the b-significant p-values, allowing every result in this paper to be independently verified against its underlying parameter choices. Tables S2–S3 and Figures S14–S16 document a complete multi-parameter sensitivity test for the Longmenshan worked example, establishing a protocol that can be applied uniformly to any other region. Together, these resources serve both as a tool for the community to evaluate the reliability of our parameter choices and as a reference benchmark for parameter selection in future b-value studies of comparable continental catalogs. To support these analyses, we developed an open-source extension to the SeismoStats package that adds adaptive-radius 2D b-value mapping with bootstrap uncertainties; together with the homogenized catalog, the parameter table, and the sensitivity-test protocol, this extension provides a comprehensive framework that the community can apply directly to other regions.

Several limitations remain. The retrospective association between low-b patches and subsequent rupture areas does not, by itself, constitute evidence of prospective forecasting skill; a formal pseudo-prospective evaluation against an independent test catalog would be required for that claim. Multi-parameter sensitivity tests have so far been carried out on the Longmenshan window as a worked example, and extensions to the remaining regions are left for future work. We hope that the homogenized catalog, the parameter table, the sensitivity-test protocol, and the SeismoStats extension released alongside this paper will, together, provide a foundation for a systematic, prospective evaluation of b-value-based precursor monitoring in mainland China.

**Acknowledgments**

This work is supported by the National Natural Science Foundation of China (No. U2239205). We thank the SeismoStats team.

## Open Research

The catalog used in this paper is available at Zhou(2026a). All codes used in this study are available at Zhou(2026b), including the advanced version of SeismoStats. The original SeismoStats can be found at https://github.com/swiss-seismological-service/SeismoStats.

## Conflict of Interest Disclosure

The authors acknowledge that no conflicts of interest have been recorded.